# DESIGN AND TEST OF WIRE-SCANNERS FOR SwissFEL

G.L.Orlandi*, M.Baldinger, H.Brands, P.Heimgartner, R.Ischebeck, A.Kammerer, F.Löhl, R.Lüscher, P.Mohanmurthy†, C.Ozkan, V.Schlott, L.Schulz, B. Rippstein, C.Seiler, S.Trovati, P.Valitutti, D.Zimoch
Paul Scherrer Institut, 5232 Villigen PSI, Switzerland,

*Abstract*

The SwissFEL light-facility will provide coherent X-rays in the wavelength region 7-0.7 nm and 0.7-0.1 nm. In SwissFEL, view-screens and wire-scanners will be used to monitor the transverse profile of a 200/10pC electron beam with a normalized emittance of 0.4/0.2 mm.mrad and a final energy of 5.8 GeV. Compared to view screens, wire-scanners offer a quasi-non-destructive monitoring of the beam transverse profile without suffering from possible micro-bunching of the electron beam. The main aspects of the design, laboratory characterization and beam-test of the SwissFEL wire-scanner prototype will be presented.

## INTRODUCTION

SwissFEL will provide coherent X-rays light in the wavelength region 7-0.7 nm and 0.7-0.1 nm [1]. Electron bunches with charge 200/10 pC and normalized emittance of 0.4/0.2 mm.mrad will be emitted by a photocathode at a repetition rate of 100Hz according to a two-bunches train structure with a temporal separation of 28ns. Thanks to a RF kicker switching the second electron bunch of the beam train into a magnetic switch-yard, the SwissFEL linac will simultaneously supply two distinct undulator chains at a repetition rate of 100Hz: the hard-Xrays line Aramis and the soft-Xrays line Athos. The electron beam will be accelerated up to 330MeV by a S-band RF Booster and to the final energy of 5.8 GeV by a C-band RF linac. Thanks to an off-crest acceleration in the RF Booster, the electron beam will experience a two-stages longitudinal compression from an initial bunch length of 3/1 ps (RMS) down to 20/3 fs (RMS) in two magnetic chicanes. The laser amplification mechanism takes advantage of a uniform distribution of the charge density along the longitudinal axis. In order to linearize the bunch-compression, two X-band RF cavities will compensate the quadratic distortion of the longitudinal phase space due to the off-crest accelerating scheme of the beam and the non-linear contribution of the magnetic dispersion. A laser-heater in the Booster section will smooth down possible micro-structures affecting the beam longitudinal profile of the beam. Macro-bunching can be detrimental to the monitoring of the beam profile based on scintillator or OTR screens (Optical Transition Radiator). This can originate phenomena of coherent radiation emission in the visible which can blind the CCD-cameras imaging the view-screen. For such a reason, as an alternative to view screens, wire-scanners (WSC) can be used to monitor the beam transverse profile. Moreover, WSCs allow a quasi-non-invasive monitoring of the beam profile which is particularly useful during FEL operations of the machine. In the following, results on design, characterization and beam test of wire-scanners for SwissFEL will be presented.

## WSC DESIGN

Wire-Scanners can be used to measure the transverse profile of the electron beam in a particle accelerator [2, 3, 4]. Carbon or metallic wires with different diameter (D) can be used to scan the beam profile with an intrinsic resolution D/4 (rms). In an electron linac, a wire stretched on a wire-fork can be vertically inserted at a constant velocity into the vacuum chamber by means of a motorized UHV linear stage. An encoder mounted on the linear stage allows the relative distance of the wire from the axis of the vacuum chamber to be measured at each machine trigger event. The interaction of the electron beam with the wire produces a "wire-signal" - scattered primary electrons and secondary particles (mainly electrons, positron and bremsstrahlung photons) - which is proportional to the number of the electrons sampled by the wire in the bunch. The Beam Synchronized Acquisition (BS-ACQ) - over a sufficient number of machine trigger shots - of the wire position and the wire-signal - detected by a loss monitors downstream the wire - allows the beam transverse profile along the horizontal or the vertical direction to be reconstructed.

In SwissFEL, view screens and WSCs will be used to monitor the transverse profile of the electron beam which varies between 500 $\mu$m and 5 $\mu$m (rms) along the entire machine. View-screens will be mainly equipped with YAG crystals. In SwissFEL, only WSCs are in principle able to discriminate the 28ns time structure of the two-bunches emitted at 100HZ by the photocathode. The SwissFEL WSCs are designed according to the following criteria, see Fig.(1): use a single UHV linear stage to scan the beam profile in the X,Y and X-Y directions; use Tungsten wire with different diameters from 5 to 13 $\mu$m to ensure a resolution in the range 1.5-3.5$\mu$m; equip each wire-scanner station with spare/different-resolution wires; detect the wire losses in the bunch charge range 10-200 pC and resolve the 28ns time two-bunches structure of the electron beam; BS-ACQ of the read-out of both the encoder wire position and the loss-monitor; wire-fork suitably designed for routine scanning of the beam profile during FEL operations; wire-fork equipped with different pin-slots where the wires can be stretched at different relative distances so that the scanning time can be minimized and optimized according to the WSC position in the machine, see Fig.(1). In SwissFEL the wire losses will be measured by means of scintillator fibers (PMMA, Poly-Methyl-Methyl-Acrylate, Saint Gobain BCF-20, emission in the green) winding up the beam pipe. The fiber are directly connected to a photomultiplier (PMT,Hamamatsu H10720-110). For more information on the detection system of the wire-signal in SwissFEL, see these conference proceedings [5].

*gianluca.orlandi@psi.ch
†presently at Massachusetts Institute of Technology, MA, USA

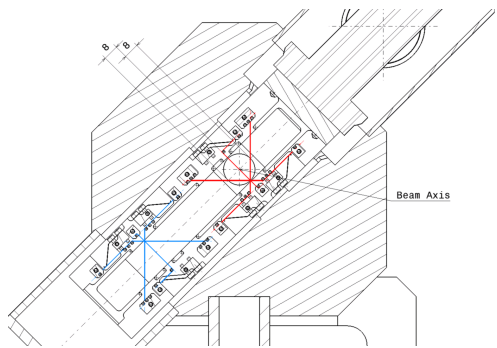

Figure 1: View of the Transverse section of the WSC vacuum chamber and, in particular, of the wire-fork and of the CF16 vacuum chamber. The wire-fork is equipped with 3 different pin-slots where the wire can be stretched. The distance of the "wire-vertex" from the vacuum-chamber axis can be set at 8, 5.5 and 3 mm in correspondence of the 3 different pin-slots.

## WSC SIMULATIONS

Most part of the primary and secondary particles resulting from the interaction of a relativistic electron beam with a metallic wire are emitted within a narrow cone with respect to the direction of incidence. Since, only a small fraction of the solid angle of emission of the wire-signal can be covered by a loss-monitor, the efficiency of the wire-losses detection strongly depends on the relative distance between the wire and loss-monitors. In order to determine the most suitable distance of a loss-monitor from a wire, the particle composition and the related energy and angular distribution of the particle shower should be evaluated. For such a purpose, Monte Carlo simulations with the FLUKA code were performed for the SwissFEL WSCs [6]. In FLUKA simulations, Tungsten wires with a diameter of $13\mu$m and $25\mu$m and beam energies of 340 MeV, 1.33 GeV, 2.99 GeV and 5.2 GeV are considered. The electron beam is modelled as a pencil beam (no energy spread). Apart from the wire, the vacuum-chamber (a CF16 stainless steel cylinder with an inner diameter of 16 mm and an outer diameter of 18mm) is the only element of the machine that is taken into account in the numerical simulation. Both the angular and the energy distributions - expressed as number of particles per unit primary particle - of the primary and secondary particles are calculated in FLUKA, see Figs.(2,3). Electrons and photons mainly contribute to the wire-signal, see Fig.(2). On the basis of the FLUKA results, 95% of the particle shower is emitted within a polar angle less than 0.1 rad, see Fig.(3). This means that, for a CF16 vacuum chamber, 95% of the wire losses intercepts the vacuum chamber within 80 mm and 4000 mm from the wire.

Beam tests of WSCs and loss-monitor response were carried out at the 250 MeV SwissFEL Injector Test Facility (SITF) [7]. For such a purpose, a scintillator fiber winding around the vacuum chamber was installed just before the high energy bending dipole of the machine [5]. The loss-monitor response was tested intercepting alternatively with several OTR screens - 300 $\mu m$ thick Si Wafer with a 200 nm Al coating - a 180 pC electron beam with energy 245 MeV and recording the corresponding time-integration of the fiber-signal read-out. The loss-monitor response measured as a function of the different OTR screen - i.e., as a function of the relative distance OTR screen loss-monitor - are shown in Fig.(4). The curve in Fig.(4) shows a minimum - i.e., a maximum of the signal - for a distance between OTR-screen and loss-monitor of about 2.5 m. This result is consistent with previous measurements carried out at SITF [8]. From the measured value of the optimum distance between screen and loss-monitor - 2.5 m - and taking into account the radius of a CF40 vacuum-chamber of SITF, a mean emission polar angle of about $\theta_{Si,300\mu m} = 7.6$ mrad can be estimated for the particle shower produced by the interaction beam-screen. In order to rescale this result to the case of a Tungsten foil, the Rossi-Greisen formula [9] can be used. The Rossi-Greisen formula allows to calculate the rms scattering angle of a charged particle experiencing multiple Coulomb while crossing a block of material. The resultant average scattering angle scales down with the charge energy and is proportional to the square-root of the normalized material thickness per radiation length of the material. Under the assumption of applicability of the Rossi-Greisen formula [9] to the considered case, the mean emission polar angle for Tungsten foils 13 $\mu m$ and 25 $\mu m$ thick can be extrapolated from the previous result $\theta_{Si,300\mu m} = 7.6$ mrad: $\theta_{Si,300\mu m}/\theta_{W,25\mu m} = 0.67$ and $\theta_{Si,300\mu m}/\theta_{W,13\mu m} = 0.93$. Taking into account the scaling formulae above, the following optimum distances of a loss-monitor from a 25 $\mu m$ and a 13 $\mu m$ thick Tungsten wire can be estimated: 0.7 m and 1 m, respectively. Under the limit of applicability of this extrapolation method, an optimum distance between wire and loss-monitor of about 1m - which is in a consistent agreement with the above reported FLUKA predictions - can be estimated.

## WSC BENCH AND BEAM TESTS

Reliable WSC measurements require a precise knowledge of the relative position of the wire with respect to the centroid of the beam at each machine trigger event. For such a purpose, the read-out of both the encoder wire-position and the loss-monitor must be acquired in a BS-

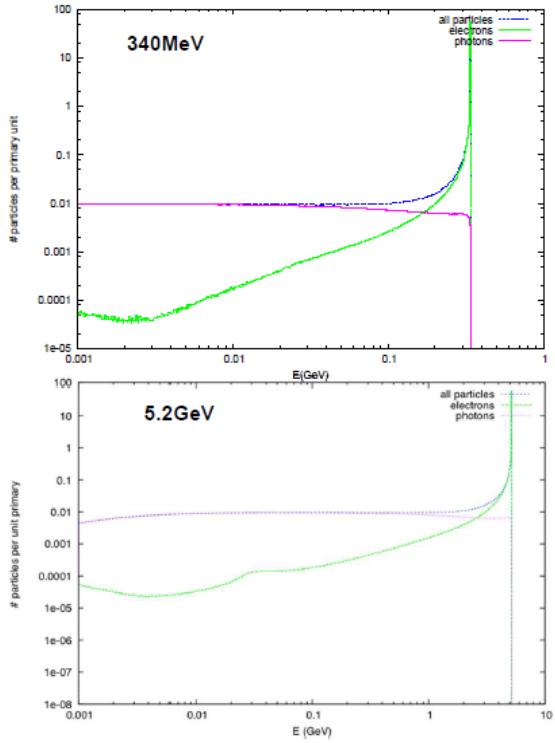
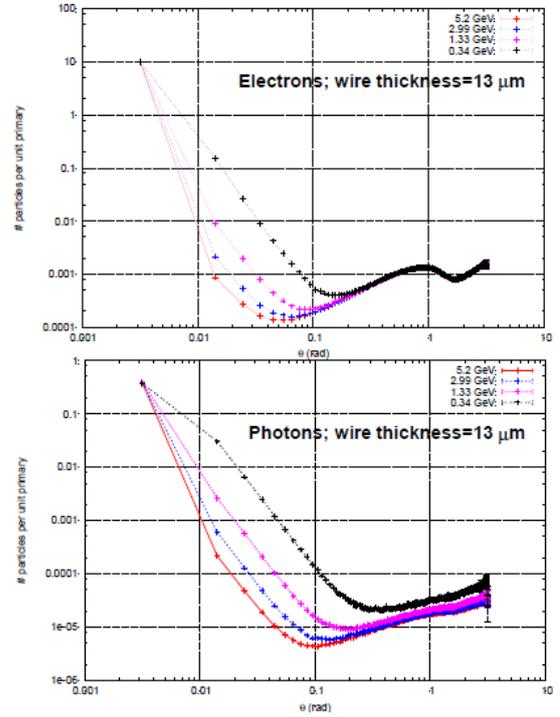

Figure 2: FLUKA results of the energy distribution of the wire losses for a beam energies of 340MeV and 5.2GeV

Figure 3: FLUKA results of the angular distribution of photons and electrons composing the wire losses

ACQ mode. The encoder read-out can provide a precise position of the wire provided that appreciable vibrations are not affecting the wire during the scan. The mechanical stability of a scanning wire can be measured in a test-bench by imaging the moving wire with a high speed camera. The wire vibration during a scan can be evaluated from the analysis of the centroid and the sigma of the projected images of the moving wire. Wire vibration measurements have been performed for different WSC set-ups, i.e., for different wire diameters, stepping motors and motor controllers. The mechanical stability of a Tungsten wire (25 um diameter) stretched on a wire-scanner fork was measured for different velocities of the stepping motor (3-phase), in the range 0.1-10 mm/s. Measurements were performed by imaging the wire by means of a high speed camera (camera frame rate 1kHz, exposure time 3us) and a microscope 10X (resolution 1px=1um). Appreciable vibrations of the wire were observed only for a wire velocity larger than 1mm/s. For a wire speed of 2mm/s, a vibration amplitude of about 0.6um (rms) was measured. A second series of vibration measurements have been performed in vacuum on a complete prototype set-up of the SwissFEL WSCs (2-phase stepping motor) for a velocity range of the wire 0.2-2 mm/s, see Fig.(1). A back-illuminated Tungsten wire (13 $\mu m$) was imaged by a camera with a frame rate of 500 fps equipped with a 200mm lens. In Figs.(5), the relative variation of the centroid and the sigma of the projected image of a 240 $\mu m$ long portion of the wire in motion at a veloc-ity of 2 mm/s are shown. After subtracting in quadrature from the rms values of the centroid and sigma distributions - Figs.(5) - the corresponding rms values for the wire at rest, it results that, for a wire velocity of 2 mm/s, the centroid vibration is about 1.3 $\mu m$ and the apparent enlargement of the sigma due to possible oscillation of the wire at a frequency higher than 500 Hz is about 0.1 $\mu m$. In conclusion, the results of the measurements of the mechanical stability of the SwissFEL WSCs indicate that, for the wire velocity range of interest for SwissFEL 0.2-2 mm/s, the measured vibration of the wire is less than the resolution limit which can be achieved in a measurement of the beam profile with a Tungsten wire with a diameter of 5 $\mu m$.

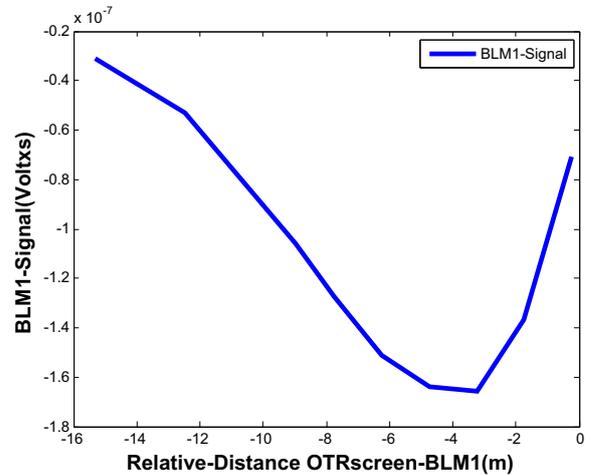

Figure 4: BLM read-out

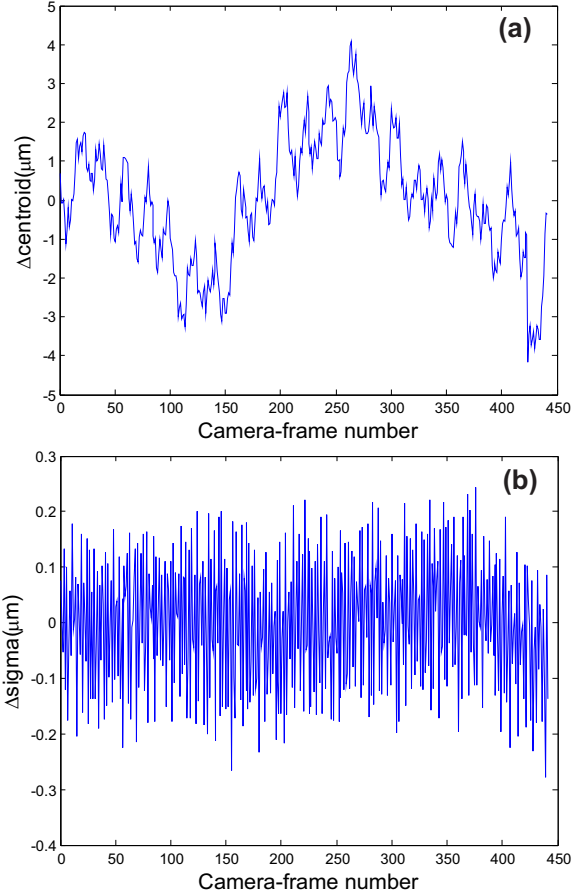

Figure 5: Wire vibration measurements results

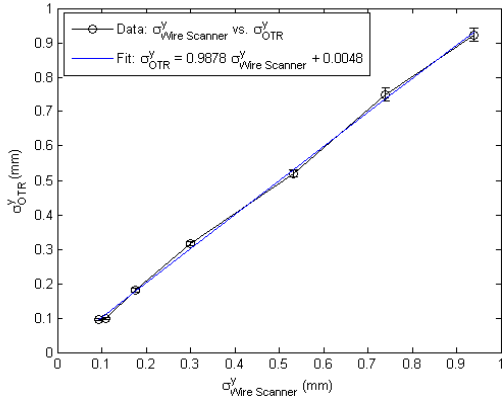

Figure 6: OTR WSC comparison

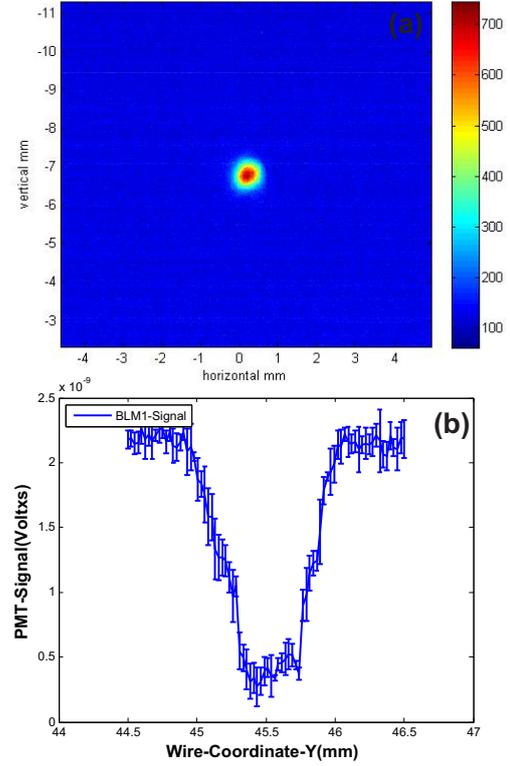

Figure 7: OTR BLM comparison

Several WSC tests on the electron beam have been also carried out in SITF. Two different techniques were adopted to detect the wire-signal produced by a Tungsten wire with a diameter of 25 $\mu m$. In the former case, the wire-signal was retrieved as the difference of the charge read-out of two Beam-Position-Monitors (BPM), the one upstream the wire and the other downstream the wire just behind the bending dipole of the high energy spectrometer. In this case a 10 Hz BS-ACQ of both the BPMs and encoder position of the wire was possible. In the latter case, the wire-signal was directly measured by a loss-monitor (scintillator fiber); no BS-ACQ available in this case. In both cases, the vertical profile of the beam measured by the WSC was compared with the beam profile measured by an OTR screen placed at the same longitudinal position of the wire. Results of both campaigns of measurements are shown in Fig.(6) (BPM read-out of the wire-signal) and in Fig.(7) (loss-monitor read-out of the wire-signal). Taking into account that the WSC data are not corrected by the transverse jitter of the beam, the OTR measurements vs. WSC measurements - see Fig.(6) - show an excellent linearity (less tha 2 % deviation) as well as the comparison of the WSC-loss-monitor and OTR measurements show a good agreement within 8%, $\sigma_Y$=0.244+/-0.002 mm (OTR), $\sigma_Y$=0.265+/-0.008 mm (WSC).

## CONCLUSIONS

Design and tests of the SwissFEL WSCs are presented. Results of the test-bench measurements of the mechanical stability of the prototype system indicate that, for the wire velocities of interest for SwissFEL (0.2-2 mm/s), the measured wire vibration is less than the intrinsic resolution expected for a 5 $\mu m$ thick Tungsten scanning the transverse profile of the electron beam. WSC tests performed on the beam at the 250MeV SwissFEL Injector Test Facility also indicate that WSC measurements of the beam transverse profile are consistent with analogous measurements performed with OTR screens.